\documentstyle[twocolumn,prb,aps,epsfig,floats]{revtex}
\begin{document}
\draft

\twocolumn[\hsize\textwidth\columnwidth\hsize\csname @twocolumnfalse\endcsname

\newcommand{\ibf}{\mbox{\boldmath $f$}}
\title{
Thermodynamics  of the anisotropic two-channel Kondo problem
}

\author{
Gergely Zar{\'a}nd$^{1,2}$, Theo Costi$^3$, Andres Jerez$^3$, 
and Natan Andrei$^4$
}
\address{
$^1$Lyman Physics Laboratory, Harvard University, Cambridge, MA \\
$^2$Research Group of the Hungarian Academy of Sciences, Institute of Physics,
TU Budapest, H-1521 \\
$^3$Institut Laue--Langevin, 6 rue Jules Horowitz, B.P. 156, 38042 Grenoble Cedex
9, France \\
$^4$Center of Materials Theory, Rutgers University, Piscataway, 
NJ 08854-0849, USA
}
\date{\today}
\maketitle
\begin{abstract}
We construct and solve numerically the thermodynamic Bethe Ansatz 
equations for the spin-anisotropic two-channel 
Kondo model in arbitrary external field $h$.
At high temperatures  the specific heat and the susceptibility
show power law dependence. For $h \rightarrow 0$ and at temperatures 
below the Kondo temperature $T_K$ 
a two-channel Kondo effect develops characterized by a Wilson ratio $8/3$, 
and a logarithmic  divergence of the susceptibility  and the linear specific 
heat coefficient. 
Finite  magnetic field, $h>0$ drives the system to a Fermi liquid fixed point 
with an unusual Wilson ratio which depends sensitively on $h$. 
\end{abstract}
\pacs{PACS numbers: 75.20.Hr, 71.10.Hf, 71.27.+a, 72.15.Qm}
]
\narrowtext

\section{Introduction}

The two-channel Kondo model (2CKM) attracted  a lot of interest  
during the past few years.\cite{TLSreviews}  This intense interest is 
mostly triggered by the rather unusual properties of this model:
The existence of finite residual entropy in zero external 
field, the vanishing of single particle scattering amplitude at 
$T=0$ temperature, and the logarithmic singularity of 
various thermodynamic quantities. All these unusual properties
appear due to the presence of a hidden and conserved {\em flavor}   
quantum number of the electrons.

Several physical systems  have been proposed to realize the 
2CKM. An unambiguous realization of the 2CKM  is  provided by 
dilute  Uranium and Cerium-based heavy fermion  compounds.\cite{TLSreviews}  
In these alloys the combination of strong spin-orbit 
interaction and crystal field symmetry effects leads to 
an effective 2CKM. 

It has been also suggested that non-commutative tunneling centers
may form two-level systems (TLS) and realize the 2CKM.\cite{VladZaw} 
In this case the localized spin 
is replaced by the position of the tunneling center, while 
the angular momenta of the conduction electrons play the role  
of the  electron spin in the 2CKM. The electron spins in the 
TLS problem play the role of silent flavor indices.
Though there have been several  concerns raised concerning the 
microscopic structure of the TLS and  the possibility of  observing
 the two-channel Kondo behavior,\cite{Altshuler_excited,Smolyarenko}
  many experiments may be consistently 
interpreted in terms of  such dynamical two-channel Kondo 
impurities.\cite{buhrman,Titanium,ZarJanZaw}

Another realization of the 2CKM is provided by
quantum dots near to their charge degeneracy point.\cite{glazman.90}
In this case  the charging states of the dot replace the impurity 
spin states and they couple to the  position variable of the 
conduction electrons. Again, the spin of the electrons acts as a 
flavor variable. Though it  appears to be extremely difficult to 
observe experimentally the non-Fermi liquid behavior associated with 
the two-channel Kondo fixed point,\cite{zarzim} some fingerprints 
of the two-channel Kondo effect have been observed 
in the capacitance of such semiconducting quantum 
dots.\cite{dots}

The above physical realizations of the 2CKM 
have the common feature that the 2CKM that describes them  is
in all cases generically strongly 
spin-anisotropic, and breaks  the full 
SU(2) spin symmetry. While this SU(2) symmetry breaking is known 
to be irrelevant at the two-channel Kondo fixed point,\cite{TLSreviews} 
it  affects both qualitatively and quantitatively the properties of the 
model at energies around and above the Kondo temperature, $T_K$,   
and has to be taken into account to make  comparison with 
experiments. 

In the present paper we generalize the Bethe Ansatz results of 
Ref.~[\onlinecite{TheoZar}] to compute both analytically and 
numerically the thermodynamics of the anisotropic
2CKM at arbitrary temperatures and magnetic fields
for various values  of the anisotropy. Below the Kondo 
temperature, $T_K$, we find that the thermodynamic properties of 
the model are very similar  to those of the isotropic 2CKM.
This observation is in full agreement with the irrelevance of the 
anisotropy at the 2CKM. However, above $T_K$ the behavior of the 
model depends crucially on spin anisotropy. We find 
that above $T_K$ all thermodynamic quantities display a 
{\em power law} behavior with an exponent determined by the 
value of the anisotropy. 
Furthermore, we find, that at finite magnetic fields the system 
flows  to an unusual Fermi liquid fixed point with an 
anisotropy-dependent Wilson ratio. 

The paper is organized as follows. In Section~\ref{sec:model}
we introduce the 2CKM. In Section~\ref{sec:BA}
we construct the thermodynamic Bethe Ansatz equations for the 
anisotropic 2CKM that we analyze in detail in Sec.~\ref{sec:thermo}.
Some details of the computation are given in the Appendix.

\section{Model}
\label{sec:model}

The anisotropic 2CKM consists of a spin $S_0 = 1/2$ impurity that
couples dynamically to the spin  of the conduction 
electrons through a strongly  anisotropic exchange interaction, 
and is described by 
the  following first  quantized   Hamiltonian ($\hbar=k_B=\mu_B=1$):\cite{VladZaw}
\begin{eqnarray}
H &=& \sum_{j = 1}^{N_e} \Bigl\{ 
-i {\partial/\partial x_j} 
+ \sum_{\alpha=x,y,z}  \delta(x_j) 
J_\alpha S_j^\alpha S_0^\alpha\bigr\}
\nonumber \\
&+& h
\Bigl(g\;S_0^z + g'\;  \sum_j S_j^z \Bigr)\;. 
\label{eq.H}
\end{eqnarray}
In this equation $x_j$ is the coordinate of the $j$'th conduction electron,
$N_e = N\times  f$ is the total number of  electrons, and 
the $S_j$'s ($j=0,..,N_e$) denote the spin 1/2   operators of 
the electrons.  Electrons also have a conserved 
flavor quantum number, $ m = \{1,..,f\}$, implicit in Eq.~(\ref{eq.H}).
In the following  we   discuss the  case $f \ge2$  as well
though for the physical realizations mentioned in the Introduction 
$f=2$. (The  $f=1$ case has been analyzed in  Refs.~[\onlinecite{WiegmanTsvelik,TheoZar}].) 

The meaning of the various terms in this Hamiltonian depends on the 
particular physical realization.  The external field $h$ couples to 
the local moment and the 
conduction electron spins with  different g-factors. 
In the TLS problem and the quantum dot case, the external 
field $h$ represents the  asymmetry of the TLS and the splitting between 
the two charging states of the quantum dot, respectively, and  $g'\equiv 0$. 
For heavy fermion systems, depending on the particular realization, 
$h$ can describe an external magnetic- or strain field, and may also 
couple to  $\sum_j S_j^z$. In the present paper we mostly study
the {\em local} susceptibility, corresponding to $g'=0$. 

In the  original formulation of the Bethe Ansatz, similar to 
the conformal field theory solution of the Kondo problem,\cite{CFT} 
the impurity spin is 'fused'  with the spin degrees of freedom of 
the conduction electrons, and the external field couples to the total 
spin, corresponding to $g'=1$. Thus the Bethe Ansatz
calculates the impurity contribution to the {\em global} 
susceptibility with $g'=g$. Lowenstein made a remarkable attempt 
to treat the $g'\ne 1$ case,\cite{Lowenstein} however,  his results 
were not fully conclusive.\cite{Lowenstein} In the present paper
we use a different strategy to cope with the $g'=0$ situation, and 
show that, similar to the case of the single channel anisotropic 
Kondo problem,\cite{Finkelst}  
the $g'=0 $ {\em local} susceptibility (i.e. the susceptibility arising
when the field couples only to the impurity spin) is simply proportional 
to the global susceptibility. Therefore, after  determining the 
appropriate normalization factor, we are  able to extract  the 
exact {\em local} susceptibility from the Bethe Ansatz calculation
with $g=g'$.  

In our analysis we shall assume that  $J_z \ne  J_x = J_y = J_\perp$
and thus the Hamiltonian has a $U(1)$ spin symmetry. This is 
true for the quantum dot and heavy fermion realizations, however, 
for the TLS  problem $J_z \gg J_x >J_y=0$.\cite{VladZaw}
Fortunately, it is not necessary to treat the general Bethe Ansatz equations
in order to determine the universal features of the TLS Hamiltonian
in the TLS case either: Under scaling the terms $J_x$ and $J_y$, describing
electron--assisted tunneling processes, become {\em rapidly}
equal\cite{VladZaw} and therefore it suffices to 
consider an equivalent effective 
model with  $J_x^{\rm eff} = J_y^{\rm eff} = 
J_\perp = J_x/2$ to determine the thermodynamics. This effective 
model has $U(1)$ orbital symmetry and can be much more easily 
analyzed by Bethe Ansatz techniques than the fully anisotropic one. 

In constructing the wave function we 
implicitly have to introduce a further electron-electron interaction
\begin{equation}
H^{el-el} = \sum_{1\le i<j\le N_e} 
 \sum_{\alpha=x,y,z} \delta(x_i-x_j)  J_\alpha S_i^\alpha S_j^\alpha\;.
\label{eq:el-el}
\end{equation}
Since electrons move with the same velocity and in the same direction,
this interaction does not modify the thermodynamics in a 
crucial way for $J_z > J_\perp$.  In particular, we have  shown 
that the specific heat 
of the  electrons is unaffected by this interaction. 
However, it does {\em rescale} the g-factor in 
Eq.~(\ref{eq.H}): $g \to g^{\rm eff}$. 
Fortunately, we shall be able to {\em compensate}
this effect  by the renormalization of $g$-factors in the 
magnetic field term. 

For $J_z < J_\perp$ the artificial electron-electron
interaction apparently generates  a  gap in the spin sector, 
and therefore it is impossible to capture the Kondo effect 
there within the algebraic Bethe Ansatz approach. 
Therefore, unfortunately, we have to restrict 
our discussion to the 
case $J_z > J_\perp$,  and our calculations are not directly 
applicable to the quantum  dot case.

In Eq.~\ref{eq.H}, only a  term $\sim h S_{0}^{z}$ has
been included, which acts as a local field in the $z$ direction. 
In reality, however, the field can also point into the 'perpendicular' 
direction corresponding to a term $\sim h_\perp S_0^x$. 
In the TLS case this term describes spontaneous tunneling between 
the TLS positions.
The general effect of the  latter term is very similar to that 
of $h S_{0}^{z}$,  but quantitatively it behaves somewhat  
differently. For a TLS, e.g., in a typical situation 
$h$ is believed to be much  larger than  $h_\perp$,\cite{Smolyarenko} 
and furthermore,  its {\em effective} value  is orders of 
magnitude  reduced  by polaronic effects.\cite{VladZaw,VladZimZaw} 
Therefore, in  most situations it is enough to include only the term 
$\sim h  S_{0}^{z}$ in the Hamiltonian.

\section{Bethe--Ansatz solution} 
\label{sec:BA}

The most important ingredients of the algebraic Bethe Ansatz (BA)  
are the various scattering matrices.  
The impurity-conduction electron S-matrix  can be constructed 
directly from (\ref{eq.H}):
\begin{equation}
R_{0j} = U^{(spin)}_{0j}(\lambda_0 - \lambda_j)_{\sigma_0 \sigma_0';
\sigma_j \sigma_j'} \otimes { Id}^{(flavor)}\;,
\end{equation}
where $\lambda_0 = -1$ and $\lambda_j = \to 0$ denote the 
``rapidities'' of the impurity and  conduction electron $j$, and 
$U$  is the $U(1)$ scattering matrix 
\begin{eqnarray}        
U_{0j}(\lambda) &=&
a(\lambda) P_{\uparrow\uparrow} + 
b(\lambda) P_{\uparrow\downarrow} +
\frac12 c(\lambda)
\bigl(S_0^+ S_j^- + S_0^- S_j^+\bigr)\;,
\nonumber \\
{a(\lambda)\over c(\lambda)} &=& 
{\sinh(i\mu + \lambda\vartheta ) \over \sinh(i\mu)}\;, 
\;\;\;  {a(\lambda)\over b(\lambda)} = 
{\sinh(i\mu + \lambda\vartheta) \over \sinh(\lambda\vartheta)}\;, 
\label{eq.S}
\end{eqnarray}
with $P_{\uparrow\uparrow}$ and $P_{\uparrow\downarrow}$ being the 
projection operators for parallel and opposite  spins.  
Excepting the small coupling regime, the connection of the 
parameters $\mu$ and $\vartheta$ 
with the bare couplings $J_z$ and $J_\perp$ is 
ambiguous\cite{WiegmanTsvelik,AndresAndrei} 
and  depends on the regularization procedure of the  Dirac 
delta and the cutoff scheme used. 
Therefore,  instead of $J_z$ and $J_\perp$, 
it is 
rather  $\mu$ and $\theta$ that should be viewed as the basic parameters
of the BA solution: $\mu$ turns out to be connected 
to the renormalized phase shift 
while the ratio  $\mu/\theta$  determines the Kondo temperature, 
below which non-Fermi-liquid correlations appear: 
\begin{equation}
T_K = (1-f {\mu\over \pi}) \;2D\exp\{ -\pi {\theta/ \mu}\}, 
\label{eq:T_K}
\end{equation}
with $D = N/L$. 

Since electrons move with the same velocity we have the  liberty 
to define their scattering matrix in a way to maintain
 integrability:
\begin{equation}
R_{ij} = U_{ij}(\lambda_i-\lambda_j) \otimes F_{ij}(\lambda_i - \lambda_j)\;.
\end{equation}
Here $U(\lambda)$ is given by Eq.~(\ref{eq.S}) and $F$ describes 
scattering in  the flavor sector:
\begin{equation}
F_{ij}(\lambda_i- \lambda_j) = {\lambda_i - \lambda_j + i c X_{ij}
\over \lambda_i - \lambda_j + i c}\;,
\end{equation}
with $X_{ij}$ the flavor exchange operator of particle $i$ and $j$ and
$c$ an arbitrary constant to be defined later. 

Starting from these scattering matrices we used the algebraic BA
 to determine the nested BA equations and then
applied the dynamical fusion  procedure of Ref.~[\onlinecite{andrei.84}] 
to eliminate
the flavor degrees of freedom. The fused equations
considerably simplify with the choice $c\equiv \frac \mu \vartheta$. 
Then the rapidities $\{\lambda_\alpha; \alpha = 1,..,M\}$
describing the spin sector of the wave function satisfy
\begin{eqnarray}
{\sinh\bigl(\mu\bigl(\lambda_\alpha + \frac i2 \bigr) + \vartheta\bigr)
\over \sinh\bigl(\mu\bigl(\lambda_\alpha - \frac i2 \bigr) + \vartheta\bigr)}
\left[ {\sinh \mu\bigl(\lambda_\alpha + i\frac f2 \bigr) 
\over \sinh\mu\bigl(\lambda_\alpha - i\frac f2 \bigr) }\right]^N
= \nonumber \\
- \prod_{\beta = 1}^M 
{\sinh \mu\bigl(\lambda_\alpha -\lambda_\beta+ i \bigr) 
\over \sinh\mu\bigl(\lambda_\alpha -\lambda_\beta- i\bigr) }\;.
\label{eq.fused}
\end{eqnarray}
The momenta of the electrons  and thus the total energy 
is determined by the  periodic boundary conditions 
\begin{equation}
e^{ik_A L f} = \prod_{\alpha=1}^M 
{\sinh \mu\bigl(\lambda_\alpha + i\frac f2 \bigr) 
\over \sinh\mu\bigl(\lambda_\alpha - i\frac f2 \bigr) }\;;\;\;\;
E =  \sum_{A=1}^N f k_A\;, \nonumber
\end{equation}
where $f k_A$ denotes the total momentum of the 
fused $f$-electron  composites, and $L$ is the system size.

In the thermodynamic limit, $L,N\to\infty$, $N/ L =D$, the 'spin
rapidities' $\lambda_\alpha$ in Eq.~(\ref{eq.fused}) are
organized into strings \cite{WiegmanTsvelik,TheoZar} of length $r$ and 
parity $v=\pm$:
$\lambda \to \{ \lambda^{(r,v)}_q\;;\;\; q = 1,..,r\}$ 
with
\begin{equation}
\lambda^{(r,v)} \leftrightarrow  \lambda^{(r,v)}_{q} = 
\lambda^{(r,v)} + \bigl[ {\textstyle \frac {r+1}2} -q\bigr] + i {\textstyle{\pi\over 4\mu}}
(1-v)\;.
\end{equation}
We have verified that to obtain a stable solution 
for $\mu< \pi/f$, $v$ and $r$ must  satisfy 
the {\em same  stability condition}  as for $f = 1$ 
\cite{Takahashi,WiegmanTsvelik}
\begin{equation}
v \sin(\mu q) \sin(\mu(r-q)) > 0\;,\;\;\;q=1,..,[r/2].
\label{eq.stability}
\end{equation}
As shown by Takahashi and Susuki,\cite{Takahashi} the 
allowed $(r,v)$ strings  can be classified  on the basis of an 
infinite (or finite) fraction  expansion of  $\mu/\pi$.  
To be specific, here we only  discuss  the simplest case 
$\mu = \pi/\nu$ and $f<\nu$, where only $\nu$ different 
stable string configurations  exist:
$n=(r,v)= (1,+), (2,+),..,(\nu-1,+)$ and $(1,-)$. 
The case $\mu = \pi/f$ represents a singular limit:\cite{TsvelikToulouse} 
For $f=1$ it corresponds to the decoupling point\cite{Emeryreview} 
while for $f=2$ it can be identified with the 
Emery-Kivelson point (see below).

\section{Thermodynamics}
\label{sec:thermo}
\subsection{Thermodynamic Bethe-Ansatz equations}
To derive the thermodynamic BA equations 
in the continuum limit 
$L,N\to \infty$ and $D\equiv N/L =cst$ 
we proceeded  in the usual way. 
We first defined the  density of  rapidities
(rapidity holes),  $\varrho_n(\lambda)$ (${\tilde \varrho}_n(\lambda)$).  
These are related to the    'excitation energies' 
$\epsilon_n(\lambda)$ through 
$\eta_n \equiv  {\tilde \varrho}_n/\varrho_n \equiv 
e^{\varepsilon_n/ T} $.
The functions $\epsilon_n(\lambda)$
are determined by the following integral equations
for $\nu > f$:  
\begin{eqnarray}
\varepsilon_{\nu}/T &=&  {g\;h\;\nu / 2T} 
- s * \ln\bigl(1+ e ^{\varepsilon_{\nu-2}/T}\bigr) + 
\delta_{\nu,f+1} \Theta(\lambda)
\nonumber \\
\varepsilon_{\nu-1}/T &=& {g \;h\;\nu  /  2T} 
+ s * \ln \bigl(1+e^{\;\normalsize \varepsilon_{\nu-2}/T}\bigr)
-\delta_{\nu,f+1} \Theta(\lambda)
\nonumber  \\
\varepsilon_j/T &=&
s * \ln\Bigl[\bigl(1+e^{\;\normalsize \varepsilon_{j+1}/T}\bigr)
\bigl(1+e^{\;\varepsilon_{j-1}/T}\bigr) \Bigr] \nonumber \\
&+& \delta_{j,\nu-2} \; s * \bigl(1+e^{-\varepsilon_{\nu}/T}\bigr)    
 -  \delta_{j,f} \Theta(\lambda)
\;\;\; (j < \nu-1).
\nonumber
\end{eqnarray}
where $s *$ denotes convolution with 
the Kernel $s(\lambda) = 1/\cosh(\pi\lambda)$, the driving term is given by
$\Theta(\lambda) = 2D/T \; {\rm arctg} ( e^{\small \pi\lambda})$   and 
$\varepsilon_0 \to -\infty$.  
The impurity contribution to the free energy is given by
$$
F^{\rm imp} = -T \int_{-\infty}^\infty
s(\lambda + {\textstyle \frac \mu \vartheta})\;
\ln\bigl(1+\exp(\varepsilon_1(\lambda)/T)\bigr)\;
d\lambda\;,
$$
and, in principle, all thermodynamic quantities can be calculated by 
taking the  derivatives of $F^{\rm imp}$.

\subsection{Analytical results}
Many of the thermodynamic properties can be determined from the asymptotic 
form of the  $\eta_n$'s. Using the Ansatz $\eta_n(\lambda\to\pm \infty) 
\approx \eta_n^\pm + b_n^\pm \;e^{\mp \pi \;\tau_\pm  \lambda}$ one obtains a 
set of algebraic equations  for the $ \eta_n^\pm$'s, $ b_n^\pm$'s and 
the exponents $\tau_\pm$. The latter exponents govern the scaling of the 
free energy in the vicinity of  the low- and high-energy fixed points
and are given by $\tau_+ = 4/(2+f)$ and $\tau_- = 2\mu/\pi$.
The crossover between the two  regimes occurs at the Kondo  scale 
Eq.~(\ref{eq:T_K}),
which emerges naturally if one rewrites the thermodynamic BA equations above 
in a 'universal' form by removing the cutoff 
$D$.\cite{Andreirev,TsvelikToulouse,remark}
The asymptotic form of the impurity free energy for $h \ll T \ll T_K$ 
is given by 
$$
-{F^{\rm imp}/ T} \sim 
\left\{
\begin{tabular}{lr}
$ S^{\rm imp} + (a + b \bigl( \frac{g h } T \bigr)^2) 
\bigl ( \frac T{T_K} \bigr)^{4/{2+f}}$ &  \phantom{nn} $f>2$, \\
$ S^{\rm imp} + \bigl(a + b \bigl(\frac {g h} T \bigr)^2\bigr)
\bigl( \frac {T} {T_K} \ln(\frac {T}{T_K} ) \bigr) $ & \phantom{nn} $f=2$,
\end{tabular}
\right.
$$
implying the divergence of the linear specific heat coefficient
$c/T$ at $h =0$ and the susceptibility as $T\to 0 $. 
The constants $a$ and $b$ above depend on the 
specific value of $f$ and $\mu$,
and the residual entropy $S^{\rm imp }$ is the same as 
in the isotropic case\cite{andrei.84}
\begin{equation}
S^{\rm imp } = \ln\Bigl[{\sin({\textstyle \frac {f\pi}{f+2}})
\over 
 \sin({\textstyle \frac \pi{f+2}})}\Bigr] \;.
\end{equation}

To determine the renormalization of the $g$-factor
and the Wilson ratio, 
we calculated the linear specific heat coefficient and the 
bulk magnetization $M^{\rm tot}_z \equiv -\partial F^{\rm tot}
/\partial h $ in the absence  of the impurity spin (but with $g'=g$).  
Similar to the case $f=1$,\cite{WiegmanTsvelik}
the linear specific heat coefficient agrees with  that of 
the spin sector of non-interacting free  electrons. 
However, the magnetization does not. 
Following similar lines as in Ref.~[\onlinecite{WiegmanTsvelik}],
the total  spin  can be related to $\eta_\nu^\pm$ and 
$\eta_{\nu-1}^\pm$ and thus the  magnetization is simply given by
\begin{equation}
M^z = -{\partial\over \partial h} 
F^{\rm tot} = g \;\langle \sum_{i=1}^{N_e} S_i^z \rangle 
= {g^2}\; { fL \over 4\pi}
{h \over 1- f \mu/\pi} \;,
\end{equation}  
with the  term, ${g^2}\; { fL / 4\pi}$,
the Pauli susceptibility of free electrons. 

From this equation it immediately follows by integration 
that at zero temperature
\begin{equation} 
F^{\rm tot} = - {1\over2} {g^2}\; { fL \over 4\pi} {h^2
\over 1- f \mu/\pi}\;.
\end{equation}
Thus the g-factor is {\em renormalized} due to the 
electron-electron interaction as 
$g \to g /(1 - \mu f/\pi)^{1/2}$,  and to compensate
the effect of Eq.~(\ref{eq:el-el}),  we have to chose 
$$
g \equiv (1 - \mu f/\pi) ^{1/2}\;.
$$

Having thus compensated the effect of the artificial 
electron-electron interaction of Eq.~(\ref{eq:el-el})
by rescaling $g$,  we can proceed to calculate 
 the  impurity contribution to the  {\em global}  susceptibility 
(defined  with $g'=1$, $g=1$ but {\em no}  electron-electron interaction).   
We find indeed that with the choice  $g \equiv (1 - \mu f/\pi)^{1/2}$ 
the low temperature (global) Wilson ratio, defined in terms of this global 
impurity  susceptibility, takes on a universal 
value
\begin{equation}
R^{\rm imp}_{\rm glob} 
= \lim_{T\to 0} \lim_{h \to 0 } {c^{\rm bulk} \over  \chi^{\rm bulk} }
{ \chi_{\rm glob}^{\rm imp}
\over  \;c_{\rm glob}^{\rm imp} } = \frac 8 3\:,
\label{eq:WRglobal}
\end{equation}
as in the isotropic case,\cite{cbulk}
proving again that exchange anisotropy is irrelevant at the 
two-channel Kondo  fixed point.\cite{PangCox,ZarJan,Ye}
In the following we shall always denote quantities that were 
calculated without (i.e. compensating) the artificial electron-electron 
interaction by the superscript 'imp'.

To capture the meaning of the parameter $\mu$ we also 
determined the impurity contribution to the 
{\em global}  susceptibility in the high 
temperature regime:
\begin{equation}
\chi_{\rm glob}^{\rm imp} = {g^2\over 4T} = \frac{1- f \mu/\pi}{4T}\;. 
\label{eq:chihighT}
\end{equation}
Using Abelian bosonization techniques we were also able to prove 
analytically 
that for  $J_\perp \ll J_z$ at high temperatures 
\begin{equation}
\chi_{\rm glob}^{\rm imp}(T\to\infty) 
= (1-2f{\delta\over \pi})^2{1\over 4T}\;,
\label{eq:ch_imp^glob}
\end{equation}
with  $\delta$ the phase shift  generated by $J_z$
 (see Appendix~\ref{app:phaseshift} for the derivation and 
the precise definition of the phase shift).\cite{VladZimZaw,ZarJan}

This immediately implies the important relation
\begin{equation}
{\mu\over \pi} = 4 {\delta \over \pi} - 4 f {\delta^2\over \pi^2}\;.
\label{eq:connect}
\end{equation}
Comparing this expression with the results of 
Ref.~[\onlinecite{VladZimZaw}] we notice immediately that $\mu/\pi$ is 
nothing but the {\em scaling dimension} of $J_\perp$, 
which satisfies the following scaling equation 
at energy scales $\omega$ well above 
$T_K$:\cite{VladZimZaw} 
\begin{equation}
{d \ln J_\perp \over d\ln (\omega_0/\omega)} 
= {\mu \over \pi}\;, \;\;\;(\omega \gg T_K)\;.
\label{eq:scaling}
\end{equation}
Here $\omega_0$ is a high energy cutoff. 
For a TLS it is  of the order of the Debye 
temperature, while for heavy fermions it is usually 
of the order of the Fermi energy.  For a quantum dot 
the cutoff is the charging energy, $\omega_0\sim E_c$. 
The effective perpendicular 
coupling $J_\perp$  at energy scale  $\omega$  can be obtained 
by simply   integrating this equation.

Eq.~(\ref{eq:connect}) is further confirmed by noticing that for $f=2$
at the Emery-Kivelson line, $\delta = \pi/4$, 
the global susceptibility  Eq.~(\ref{eq:chihighT}) identically vanishes, 
in complete agreement with the results of 
Refs.~[\onlinecite{EmeryKivelson}].
The point $\delta = \pi/2f$ corresponding 
to $\mu = \pi/f$ is highly singular, 
and   needs special care:\cite{TsvelikToulouse} At this particular
 point the amplitude  of the leading irrelevant operator, 
responsible for the divergence  of the   susceptibility 
and the linear specific  heat  coefficient, becomes 
zero.\cite{EmeryKivelson} 

The global susceptibility, $\chi_{\rm glob}^{\rm imp}$, and  the 
associated global Wilson ratio, $R^{\rm imp}_{\rm glob}$, defined
in Eq.~\ref{eq:WRglobal}, are useful for magnetic  Kondo 
systems.  However, for quantum dots and TLS's 
it is rather the {\em local } impurity susceptibility 
that is of interest, i.e., the response of the system 
to an external field coupled only to the impurity spin.
Hence, we studied  the impurity  contribution to the susceptibility 
when $g=1$ and $g'=0$ (and in the absence  of the artificial 
electron-electron interaction). 
In order to determine this we generalized the path integral 
derivation of Ref.~[\onlinecite{Finkelst}] to show that 
\begin{equation}
F^{\rm imp}(h, T, g, g'=g)  = 
F^{\rm imp}(\tilde h, T, g,g'=0)\;,
\label{eq:finkel}
\end{equation} 
where $\tilde h = h (1 - 2f\delta/\pi) = h (1-\mu\;f/\pi)^{1/2}$. 
To  establish the second equality we used Eq.~(\ref{eq:connect}). 
Thus we can calculate the local impurity properties (those for $g=1$, $g'=0$
and no electron-electron interaction) from the BA equations 
with  $g' = g= (1-f \mu/\pi)^{1/2}$  
by simply rescaling the field  $h$ 
according to Eq.~(\ref{eq:finkel}).\cite{miracle}

Henceforth, unless otherwise stated, we shall denote the
local impurity susceptibility obtained from the BA solution in combination
with Eq.~(\ref{eq:finkel}) by $\chi^{\rm imp}$ and the associated 
local Wilson ratio by $R^{\rm imp}$ (see below). 

Following the above procedure we find 
at high temperature
\begin{eqnarray}
&& \chi^{\rm imp}  = {1 \over  4T}\Bigl[1 -B\bigl( {T_K \over T}\bigr)^
{ 2 \mu/\pi }\Bigr]\;, \\
&& C^{\rm imp}  \sim 
\left(T_K \over  T\right)^{ 2 \mu/\pi } \;.
\label{eq:HighTemp}
\end{eqnarray}
Note that at high temperatures the specific heat exhibits 
a {\em power law} behavior which crosses over to a logarithmic 
behavior in the isotropic case $\mu\to 0$. Similarly, the corrections to
the susceptibility about the free behavior are power law
like and these power laws give way to logarithmic corrections in the 
isotropic limit  $\mu\to 0$. We note that the above exponent $2\mu/\pi$
is formally the same as found for the $f=1$ case \cite{TheoZar}. 
However, the relation between $\mu$ and the bare couplings is quite 
different in the two cases and involves the channel number 
$f$ (see Eq.~(\ref{eq:connect})).

The power-law corrections can be very easily understood from the 
scaling picture. Expanding the free energy in terms of 
$J_\perp$ one finds that the leading correction is second order 
in $J_\perp$. Making use of the scaling equation Eq.~(\ref{eq:scaling}) 
it immediately follows that the leading corrections to the free energy 
behave as $T^{1-2\mu/\pi}$, implying Eq.~(\ref{eq:HighTemp}). 
Similar arguments lead to the conclusion 
that the impurity-induced resistivity correction behaves 
at high temperatures as
\begin{equation} 
\rho^{\rm imp} = A + B \left(T_K \over  T\right)^{ 2 \mu/\pi }\;
\end{equation}
for $T\gg T_K$.

At low temperatures, $T\ll T_{K}$, some care is required in discussing
thermodynamic properties. In contrast to the case 
$f=1$, where $F$ is analytical around $T=h=0$ \cite{WiegmanTsvelik} 
here  the $h = 0$, $T=0$  point is an essential singularity and 
the two limits $T\rightarrow 0$ and $h \rightarrow 0$
are not interchangeable. Taking the limit $h \rightarrow 0$ first,
we find the following non-Fermi liquid behavior
\begin{eqnarray}
\chi^{\rm imp}
 &\sim &  {1\over T_K} \ln({T\over T_K}) \;,\\ 
C^{\rm imp}  & \sim & {T\over T_K} \; \ln({T\over T_K}) \; , \\
\tilde{R}^{\rm imp}
& = & \lim_{T\to 0} \lim_{h \to 0 } {c^{\rm bulk}
\over \chi^{\rm bulk}}
{\chi^{\rm imp}
\over  C^{\rm imp} } = {1\over 1- f {\mu/ \pi}} \frac 8 3\:.
\label{eq:WRlimit1}
\end{eqnarray}
Here, we have defined a local Wilson ratio, 
$\tilde{R}^{\rm imp}$. It is expressed 
in terms of the local impurity susceptibility $\chi^{\rm imp}$, and 
the impurity contribution to the specific heat $c^{\rm imp}$.
It differs from the Wilson ratio defined in Eq.~\ref{eq:WRglobal} (which
is the usual definition for this quantity in the magnetic Kondo problem)
by having $g'=0$. We see that this local Wilson ratio depends on anisotropy
$\mu$ and thus the phase shift $\delta$, and is therefore {\em not} universal.

We now consider the limit of $T\rightarrow 0$, with $h$
remaining either finite, or taken to zero subsequently. In this case, 
and for $h \ll T_{K}$, we find from the numerical results of the 
next section, that there is a low energy scale $T_{\rm FL}=h^{2}/T_{K}$, 
below  which the thermodynamics is Fermi liquid like. In particular
for $T\ll T_{\rm FL}$ and small magnetic fields $h \ll T_{K}$ we find 
\begin{eqnarray}
\chi^{\rm imp}(h,T\ll T_{\rm FL})
 & \sim &  - \ln(h /T_{K}) \;,\\ 
C^{\rm imp}(h ,T\ll T_{\rm FL})  & \sim  &{T\over T_F}\;,
\label{eq:Thermolimit2}
\end{eqnarray}
with $T_{\rm FL}\sim h^{2}/T_{K}$ (for $h \ll T_{K}$).
A local Wilson ratio for arbitrary local magnetic field $h$ can be
defined as
\begin{eqnarray}
R^{\rm imp}(h)
& = & \lim_{T\to 0}{c^{\rm bulk}(h)
\over \chi^{\rm bulk}(h)}
{\chi^{\rm imp}(h)
\over  C^{\rm imp}(h) }\:.
\label{eq:ThermoLimit2}
\end{eqnarray}
In contrast to the $f=1$ case, for which this quantity is independent of 
$h$, but dependent on anisotropy 
(being given by $R^{\rm imp, f=1}(h) =
2/(1-\mu/\pi)$ \cite{TheoZar,note-global}), for the present $f=2$ 
case it depends explicitly on {\em both} $h$ and anisotropy 
(Fig.~\ref{fig6}). 
This important result, which is consistent with the result for the 
isotropic case \cite{SacramentoSchlottmann}, will be discussed in the 
following section on the numerical solution.
We note here, however, that the local Wilson ratio for the $f=2$ case 
agrees with the $f=1$ local Wilson ratio in the case of asymptotically 
large magnetic fields, $h \gg T_K$, i.e. in this case we have
\begin{equation}
R^{\rm imp}(h\gg T_K) = {2 \over {1-{\mu\over \pi}}}\;,
\end{equation}
although the meaning of $\mu$ is  different in the two 
cases (see Eq.~(\ref{eq:connect})).
The detailed dependence of this local Wilson ratio on $h$ 
will be discussed in the next section.

\begin{figure}
\begin{center}
\psfig{figure=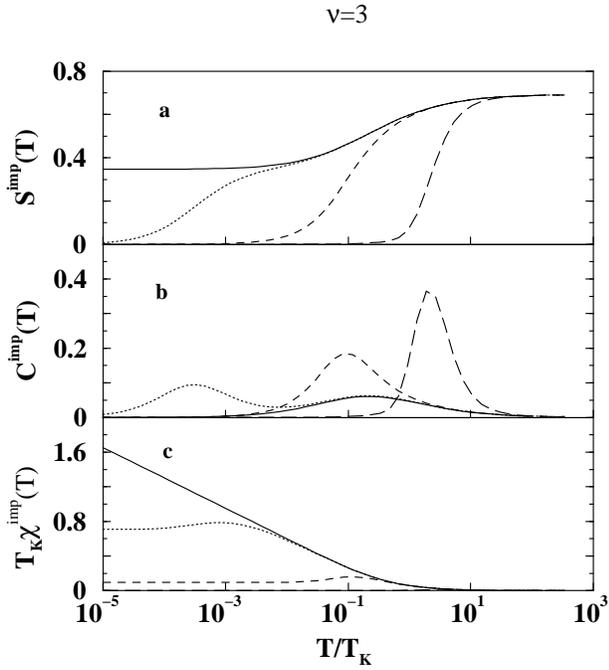,width=8.0cm}
\end{center}
\vspace*{0.05cm}
\caption{
The impurity contribution to the entropy, $S^{\rm imp}$, specific heat, 
$C^{\rm imp}$, and (local) susceptibility, $\chi^{\rm imp}$, as 
functions of $T/T_{K}$ for magnetic fields
$h/T_{K}=2^{-6}$ (solid),
$h/T_{K}=2^{-4}$ (dotted), $h/T_{K}=1$ (dashed) and $h/T_{K}=2^{4}$  
(long-dashed) for the case
$\nu=3$ corresponding to the largest anisotropy studied. 
}
\label{fig1}
\end{figure}

\subsection{Numerical solution}
In order to obtain the thermodynamics at all temperatures, it was 
necessary to solve the thermodynamic BA equations of Sec.IVA 
numerically. A procedure for
doing this, which is valid for arbitrary values of the magnetic field
$h$ and temperature has been developed in
Ref.~[\onlinecite{TheoZar}] for $f=1$. With small modifications, the
same procedure applies also to the present case. We considered
anisotropies given by $\mu/\pi=1/\nu$ with $\nu=3,4,5,6$. 
In Fig.~\ref{fig1}a-c we show the thermodynamics of the anisotropic 2CKM 
for a  large anisotropy ($\nu=3$) as one
might have in a realistic system. The characteristic non-Fermi
liquid behavior, in particular the $\ln(2)/2$ entropy \cite{andrei.84} 
and the logarithmically divergent $\chi^{\rm imp}(T)$ and 
$C^{\rm imp}(T)/T$, are found at zero field. A finite field, 
$h>0$, restores Fermi liquid behavior at temperatures below a low
energy scale $T_{\rm FL}=h^{2}/T_K$, as found for the isotropic,
$\nu=\infty$, case\cite{SacramentoSchlottmann}. The non-Fermi liquid 
behavior for $0<h<T_{K}$ is therefore restricted to an intermediate
range of temperatures, $T_{\rm FL} < T < T_K$, and we see that a
clear signature of such behavior (such as the two peaks
in the specific heat with each peak having only $\ln(2)/2$ entropy,
or a $\ln(T)$ behavior of $\chi(T)$ for a temperature range below
$T_K$) is possible, even at moderate magnetic fields, $h\sim T_{K}/16$.

\begin{figure}
\begin{center}
\psfig{figure=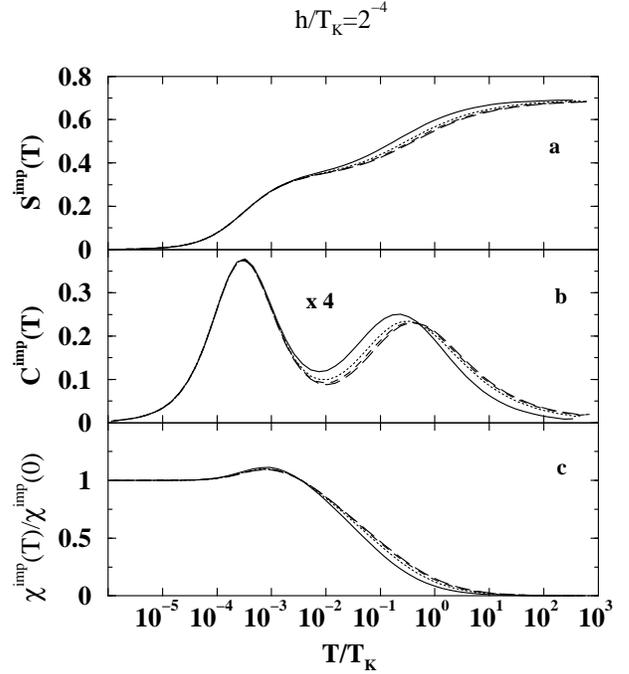,width=8.0cm}
\end{center}
\vspace*{0.05cm}
\caption{
The anisotropy dependence of the entropy, specific heat 
and susceptibility at small magnetic fields ($h/T_K=2^{-4}$), 
(note that (b) is scaled by a factor 4 for comparison with the corresponding
case of large magnetic fields shown in Fig. \ref{fig3}). 
The anisotropies shown are for $\mu/\pi=1/\nu$ 
with $\nu=3$ (solid), $\nu=4$ (dotted), $\nu=5$ (dashed) and $\nu=6$ 
(long-dashed)
}
\label{fig2}
\end{figure}
\begin{figure}
\begin{center}
\psfig{figure=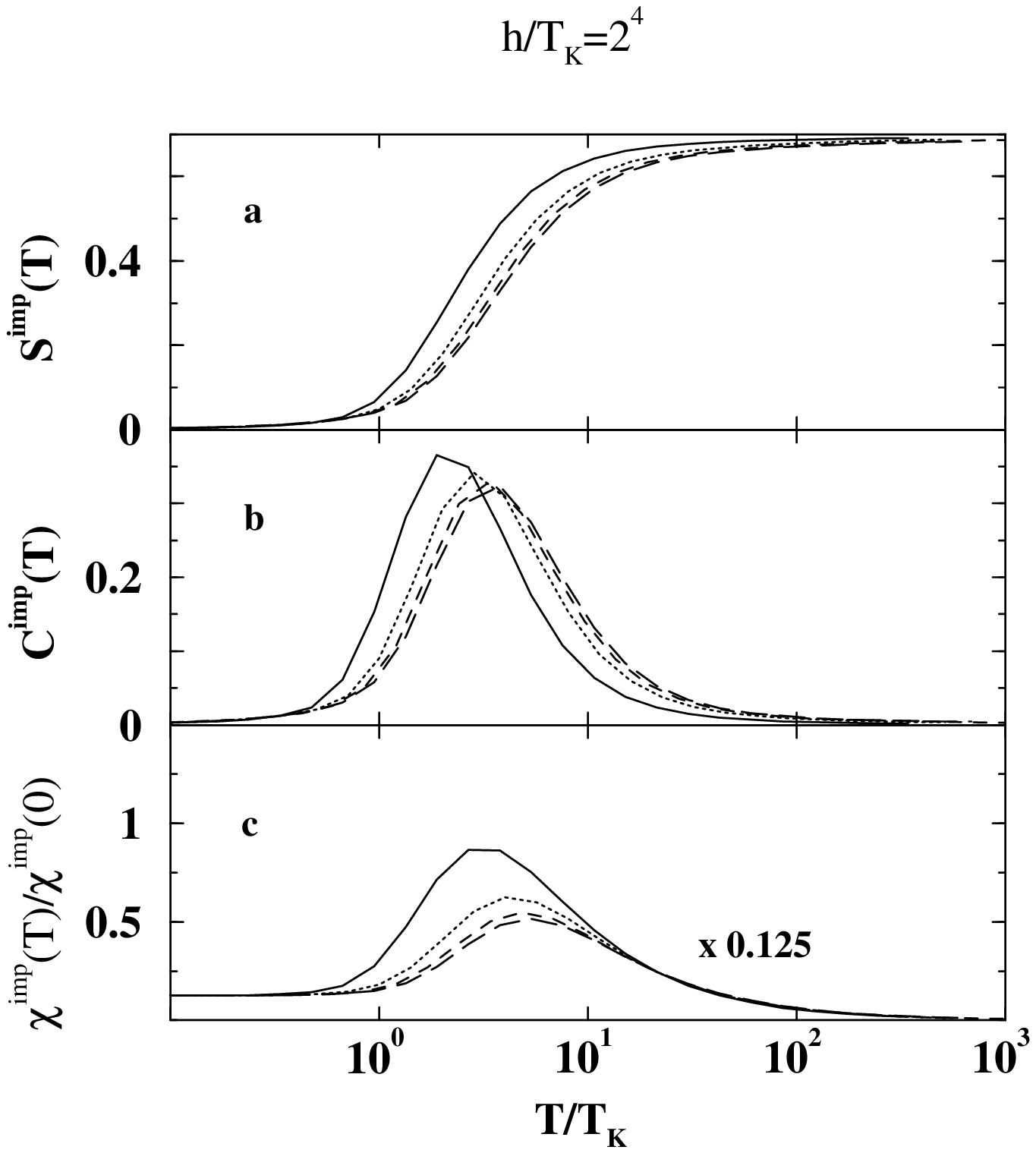,width=8.0cm}
\end{center}
\vspace*{0.05cm}
\caption{
The anisotropy dependence of the entropy, specific heat 
and susceptibility at large magnetic fields ($h/T_K=2^{4}$), 
(note that (c) is scaled by a factor 0.125 for comparison with
the corresponding case for small  magnetic fields shown in 
Fig. \ref{fig2} ). The anisotropies shown are for $\mu/\pi=1/\nu$ 
with $\nu=3$ (solid), $\nu=4$ (dotted), $\nu=5$ (dashed) and $\nu=6$ 
(long-dashed)
}
\label{fig3}
\end{figure}

\begin{figure}
\begin{center}
\psfig{figure=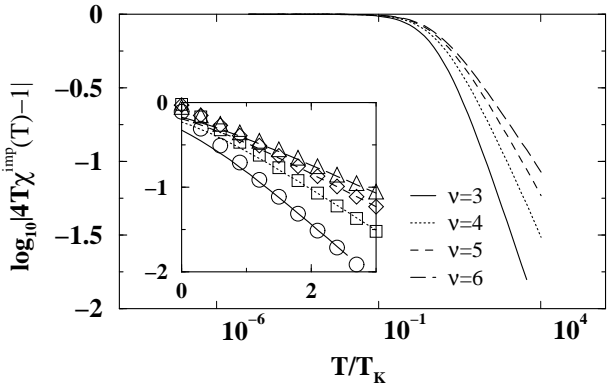,width=8.0cm}
\end{center}
\vspace*{0.05cm}
\caption{
The susceptibility at high temperatures, 
$T\gg T_K$, for anisotropies $\mu/\pi=1/\nu$ and $\nu=3,4,5,6$. The 
impurity  susceptibility at $T\gg T_{K}$ has the form
$\chi^{\rm imp}(T)=\frac{1}{T}(1/4 - B(T_K/T)^{2\mu/\pi})$, 
i.e. the corrections to the free behavior are power laws with exponents 
$2\mu/\pi=2/3,1/2,2/5,1/3$ for $\nu=3,4,5,6$. This is illustrated
in the inset, which shows $\log4(T\chi^{\rm imp}(T)-1)$ versus
$\log(T/T_{K})$. Straight lines with slopes
$-2\mu/\pi=-1/3,-2/5,-1/2,-2/3$ for $\nu=3,4,5,6$ are indicated
by symbols and are well reproduced by the numerical results.
}
\label{fig4}
\end{figure}
\begin{figure}
\begin{center}
\psfig{figure=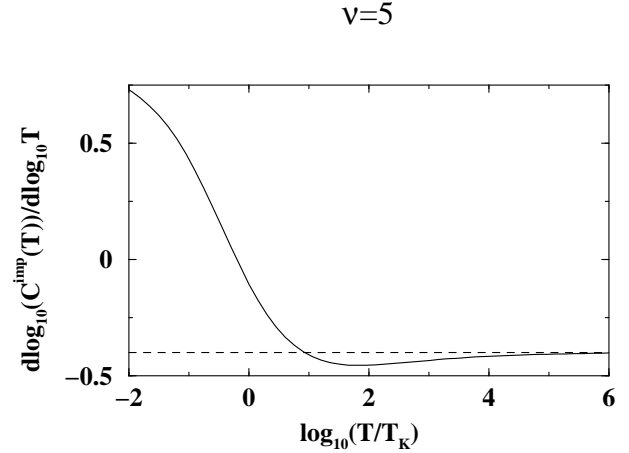,width=8.0cm}
\end{center}
\vspace*{0.05cm}
\caption{
The specific heat at high temperatures, 
$T\gg T_K$, shows power law behavior $C^{\rm imp}(T)\sim (T_K/T)^{2\mu/\pi}$
with $\mu/\pi=1/\nu$. The logarithmic derivative of this, 
$d\log\;C^{\rm imp}(T)/d\log\;T$, (solid line) approaches 
$-2\mu/\pi$ at $T\gg T_{K}$ and is shown here for 
$\nu=5\;$ ($\mu/\pi=1/5$). It is seen to approach
$-2\mu/\pi=-2/5$ (dashed line) at high temperatures. 
}
\label{fig5}
\end{figure}

In Fig.~\ref{fig2}a-c and Fig.~\ref{fig3}a-c  we show the effect of 
different anisotropies on the thermodynamics for both a small 
external fields ($h \ll T_{K}$, Fig.~\ref{fig2})
and a large magnetic fields  ($h \gg T_{K}$, Fig.~\ref{fig3}). 
At $h \ll T_{K}$, the main effect of anisotropy is to modify 
the thermodynamics at intermediate, $T_{\rm FL}\lesssim T\lesssim T_{K}$, 
and high temperatures, $T > T_{K}$. For large magnetic field, $h\gg T_{K}$,
the anisotropy modifies the thermodynamic properties at temperatures, 
$T\gtrsim T_{K}$ (Fig.~\ref{fig3}).

A characteristic feature of the present model ($f=2$), 
is that, similar to the case dissipative two state systems,\cite{TheoZar}  
it has  {\em power-law} corrections in its thermodynamics at high 
temperatures ($T\gg T_{K}$).  The exponents are uniquely
related to the anisotropy parameter $\mu$ (cf. Eq.~\ref{eq:HighTemp}).
This is in contrast to the corresponding isotropic models which
have logarithmic corrections at high temperatures. 
Fig.~\ref{fig4} and Fig.~\ref{fig5} show this for the susceptibility 
and specific heat, respectively. 

\begin{figure}
\begin{center}
\psfig{figure=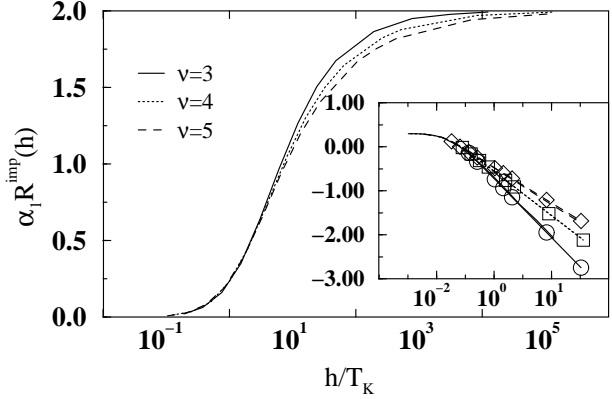,width=8.0cm}
\end{center}
\vspace*{0.05cm}
\caption{
The magnetic field ($h$) dependence of the (local) Wilson ratio, 
$R^{\rm imp}(h)$ for different anisotropies 
$\mu/\pi=1/\nu$ ($\nu=3$ (solid), $\nu=4$ (dotted), $\nu=5$ 
(dashed)). At $h\gg T_K$,
$\alpha_{1}R^{\rm  imp}(h) =
2 - a (T_{K}/h)^{2\mu/\pi}$.
This is illustrated in the inset, which shows 
$\log(2-R^{\rm imp}(h))$ versus $\log(h/T_{K})$.
Straight lines with slopes $-2\mu/\pi=-2/3,-1/2,-2/5$, for
$\nu=3,4,5$, are indicated by symbols and are well reproduced by 
the numerical results.
}
\label{fig6}
\end{figure}

Fig.~\ref{fig6} shows the remarkable magnetic field dependence of the local 
 Wilson ratio, $R^{\rm imp}(h)$. This magnetic field dependence is 
consistent with the result for the isotropic case 
$\mu\rightarrow 0$.\cite{SacramentoSchlottmann} It is quite unexpected 
for the following
reasons. First, the corresponding local Wilson ratio 
for  the $f=1$ case discussed in 
Refs.~[\onlinecite{WiegmanTsvelik,TheoZar,sassetti.90}])
is independent of the magnetic field  and is given by
$$
R^{\rm imp, f=1}={2 \over \alpha_{1}}={2 \over (1-\mu/\pi)}.
$$
It depends only on the anisotropy $\mu$ (that corresponds to the
dissipation strength $\alpha_1$ in the equivalent dissipative 
two-state system \cite{TheoZar}).
In contrast, for the present case ($f=2$), the local Wilson ratio depends 
both on the anisotropy $\mu$ and on the magnetic field $h$.
Thus, even though the regime $T\ll T_{\rm FL}$ describes 
a Fermi liquid, in the sense of Eq.\ref{eq:ThermoLimit2}, the Fermi
liquid state appears different to that for the $f=1$ case. We note that
the local Wilson ratio for $f=2$ deviates increasingly from the
usual $f=1$ Fermi-liquid Wilson ratio with decreasing  $h$, i.e.,
as the range over which non-Fermi liquid behavior dominates 
increases. The
numerical results for  $R^{\rm imp}(h)$ in Fig.~\ref{fig6} indicate
that $R^{\rm imp}(h)$ vanishes as $h\to 0$.
This result is consistent  with the numerical analysis of 
the $h\to 0$ estimates of the 
susceptibility and specific heat in Eq.~\ref{eq:Thermolimit2}. 
In contrast, from Eq.~\ref{eq:WRlimit1}
$\tilde{R}^{\rm imp}(T\to 0,h = 0) $ is finite, so we see again that
the two limits $T\to 0,h=0$ and $T=0,h\to 0$ cannot be exchanged.
The inset to  Fig.~\ref{fig6} shows that $R^{\rm imp}(h)$ exhibits power
law behavior at large magnetic fields. For $h\gg T_{K}$ we find
$$\alpha_{1}{R}^{\rm imp} = 2 - a (T_{K}/h)^{2\mu/\pi}.$$ 

\begin{figure}
\begin{center}
\psfig{figure=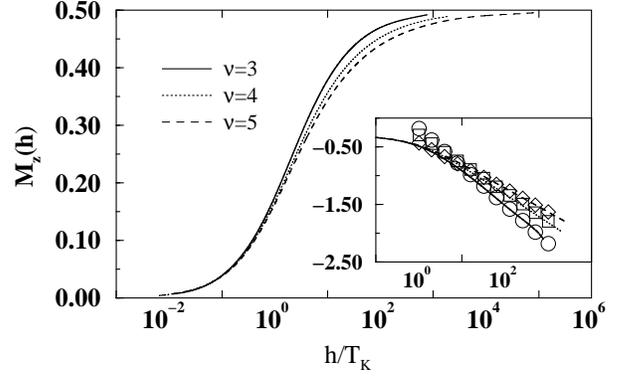,width=8.0cm}
\end{center}
\vspace*{0.05cm}
\caption{
The magnetic field ($h$) dependence of the impurity magnetization, 
$M_{z}(h,T=0)$, for different anisotropies $\mu/\pi=1/\nu$
($\nu=3$ (solid), $\nu=4$ (dotted), $\nu=5$ 
(dashed)). At $h\gg T_K$,
$M_{z}(h) =1/2 - b (T_{K}/h)^{2\mu/\pi}$.
This is illustrated in the inset, which shows
$\log(1/2-M_{z}(h))$ versus $\log(h/T_{K})$.
Straight lines with slopes $-2\mu/\pi=-2/3,-1/2,-2/5$, for 
$\nu=3,4,5$, are indicated by symbols and are well reproduced
by the numerical results.
}
\label{fig7}
\end{figure}

Finally, for completeness, we show in 
Fig.~\ref{fig7} the impurity magnetization (or polarizability) 
in the presence of a local field, 
$ M_{z}(h) =  \langle S_0^z\rangle (h)$. For each 
anisotropy, $\mu$, $M_{z}(h)$ is a universal function of $h/T_{K}$ 
(the same holds for $R^{\rm imp}(h)$). The approach 
of $M_{z}(h)$ to the free value for $h \gg T_{K}$ depends on 
anisotropy and is found numerically to behave like 
$$
M_{z}(h)={1/2} -b(T_{K}/h)^{2\mu/\pi}.
$$ 
We expect that 
in the isotropic limit $\mu\rightarrow 0$ the above power laws 
will give way to logarithmic corrections in the same way that
the power law corrections to the high temperature thermodynamics
gave rise to logarithmic corrections in this limit.

\section{Conclusions}
\label{sec:concl}

In summary, we presented a detailed analysis of the 
thermodynamics  of spin-anisotropic 
2-channel Kondo model by using the Bethe Ansatz 
technique combined with bosonization and renormalization group
arguments, and discussed quantitatively the role  of the anisotropy and
the magnetic field. 

We showed that at  high temperatures 
the thermodynamics is very different from that of the 
isotropic model: 
The local  impurity susceptibility is essentially 
free impurity-like, however, the coefficient of the 
global susceptibility is non-universal, and is related to 
the phase shift $\delta$ generated by the coupling $J_z$. 
In particular,  the global susceptibility vanishes at the 
Emery-Kivelson line, $\delta = \pi/2f$.  More interestingly, 
the impurity specific  heat (and the corrections to the susceptibility
about the free behavior) exhibits a power law behavior at 
high temperature  
\begin{equation}  
C^{\rm imp}(T\gg T_K,h)   \sim  
\left(T_K \over  T\right)^{ 2 \mu/\pi } \;.
\end{equation}
The anomalous exponent $\mu$ is the anisotropy parameter  
in the Bethe Ansatz. We have shown that  $\mu/\pi$ is just the 
anomalous scaling exponent of the spin flip term $J_\perp$
at high temperature  and that it is related to  the phase 
shifts generated by $J_z$
through\cite{VladZimZaw} 
\begin{equation}
{\mu\over \pi} = 4 {\delta \over \pi} - 4 f {\delta^2\over \pi^2}\;.
\end{equation}

On general scaling arguments,\cite{VladZaw,VladZimZaw}
a similar power-law dependence is expected to appear in the  
impurity resistivity. For $f=2$  we find 
\begin{equation} 
\rho^{\rm imp}(T\gg T_K) \sim \left( {T_{K}\over T} \right) ^{2 \mu/\pi}\;.
\end{equation} 
rather then a  simple logarithmic scaling. 

For $h=0$, and for temperatures below the Kondo temperature,
the  thermodynamics is governed by the isotropic 2CK fixed point and most of 
the thermodynamic properties resemble very much those
of the fully isotropic model, even for strong anisotropies. 

For finite $h$  
we showed that the non-Fermi liquid behavior found for $h=0$
persists for an intermediate region of temperatures $T_{\rm FL}< T < T_{K}$,
provided $h<T_{K}$ so that the new scale $T_{\rm FL}\sim h^2/T_K$ 
is well below $T_{K}$, just as in the isotropic 
case.\cite{SacramentoSchlottmann}
We also showed that the Fermi liquid behavior below $T_{\rm FL}$ is unusual
in that the  Wilson ratio, $R^{\rm imp}(h)$, 
depended very sensitively on the magnetic field $h$.  
(in contrast to the $f=1$ case, which is completely independent of $h$) 
and we calculated the detailed dependence of this quantity for 
several anisotropies. 

While here we focused our attention to specific values of the 
anisotropy, our calculations can be easily generalized to 
other anisotropy values and serve as a basis for any 
interpolation necessary for cases where there is wide 
distribution of anisotropies present.

The authors are grateful to A.M. Tsvelik and A. Zawadowski for
useful discussions. This research has been supported by 
the Hungarian Grants OTKA T026327, OTKA  F030041, and
OTKA T029813, and  NSF grant No. DMR 99-81283. The hospitality
of the Institut-Laue Langevin during a visit (GZ) is gratefully
.

\appendix

\section{Derivation of Eq.~(17)}
\label{app:phaseshift}
To prove Eq.~(\ref{eq:ch_imp^glob}) let us consider the limit 
$J_\perp\to0$ of Eq.~(\ref{eq.H}). In this case the 
interaction part of the Hamiltonian becomes: 
\begin{equation}
H_{\rm int} = {J_z\over 2} \sum_{j = 1}^f
(\psi^\dagger_{\uparrow,j}(0)\psi_{\uparrow,j}(0) - 
\psi^\dagger_{\downarrow,j}(0)\psi_{\downarrow,j}(0)) S^z_0\;,
\end{equation}
where the electronic field operator $\psi^\dagger_{\alpha,j}$
creates chiral (right-moving)  electron with  spin 
$\alpha = \{+,-\} = \{\uparrow,\downarrow\}$ and channel index
$j = \{1,..,f\}$. Thus the interaction simply produces a
spin-dependent potential scattering, and gives rise to a 
phase shift $\delta$
\begin{equation}
\psi_{\alpha,j}(x = 0^+) = \psi_{\alpha,j}(x = 0^-) 
e^{ -  4i\alpha \delta S_0^z }\;.  
\label{eq:phase}
\end{equation}

In general, the connection between $J_z$ and $\delta$ depends on the 
particular cutoff scheme used except for the small coupling limit, 
$J_z \ll 1$. To be specific, here we shall use Abelian bosonization 
on a system of finite size $L$ and the cutoff-scheme associated 
with it.\cite{Jan:boso}  
In the bosonization procedure we rewrite the Hamiltonian  as 
($g=g'=1$):
\begin{eqnarray}
H & = & \sum_{\alpha,j} \int {dx\over 4\pi} (\partial_x \Phi_{\alpha,j})^2 
+ {2\pi \over L} {1\over 2}  \sum_{j,\alpha} N_{\alpha,j}^2  
\nonumber \\
& + &  {J_z\over 2} S_0^z 
\sum_{\alpha,j} \bigl( {\alpha\over 2\pi} \partial_x \Phi_{\alpha,j}+ {1\over L} \alpha N_{\alpha,j} \bigr)  
\nonumber \\
& + & h  \bigl( \sum_{\alpha,j} {\alpha \over 2}  N_{\alpha,j} 
+ S_0^z\bigr)\;,
\end{eqnarray}
where the external field $h$ couples to the {\em total} 
pseudospin of the system,  $N_{\alpha,j}$ denotes 
the total number of electrons  with respect to the ground state 
with  spin $\alpha$  in channel $j$, and the free 
bosonic fields satisfy:
\begin{equation}
[\partial_x \Phi_{\alpha,j}(x),  \Phi_{\alpha',j'}(x')] = 
- i\; 2 \pi \;\delta_{jj'} 
\delta_{\alpha \alpha'} \delta(x-x')\;.
\end{equation}
The original fermion fields can be represented as
\begin{equation} 
\psi_{\alpha,j}(x)  = {1\over \sqrt{a} }\;F_{\alpha,j}\; 
e^{- i \Phi_{\alpha,j}(x)}\;,
\label{eq:fermionize}
\end{equation}
where $F_{\alpha,j}$ denotes the Klein factor, and $a$ is a small 
distance cut-off of the order of the lattice spacing. 

The phase shift can be most easily calculated by introducing 
charge and spin  fields and quantum numbers: 
\begin{eqnarray} 
\label{2ck:transformationN}
&& {\textstyle
\left( \begin{array}{c}
        {\Phi}_{c,j} \\   {\Phi}_{s,j} 
         \end{array}
\right)}
\equiv
{1\over \sqrt{2}} 
{\textstyle     
        \left( \begin{array}{cc}
        1 & \phantom{-}1 \\
        1 & -1 
        \end{array} \right)
\left( \begin{array}{c}
         \Phi_{\uparrow,j} \\  \Phi_{\downarrow,j} 
        \end{array} \right) \; ,
}
\\
&& N_{c,j} \equiv N_{\uparrow,j} + N_{\downarrow,j} \;,
\\
&& N_{s,j} \equiv {1\over 2} \bigl( N_{\uparrow,j} - N_{\downarrow,j}\bigr) 
\;,
\end{eqnarray}
and  performing a  unitary transformation  on the Hamiltonian
by $U = e^{i \sum_j J_z 
\Phi_{s,j}(0)  S^z_0 /(2\pi\sqrt{2})}$, resulting in 
\begin{equation}
\partial_x \Phi_{s,j} (x)  \to \partial_x\Phi_{s,j}(x)  - 
{J_z\over \sqrt{2}} S_0^z \delta(x)\;,
\label{eq:shift}
\end{equation}
and the  'noninteracting'  Hamiltonian:
\begin{eqnarray}
H & = & H_0 + H_s\;, \\
H_0  & = & \sum_{j=1}^f \sum_{\mu= c,s} \int {dx\over 4\pi} 
(\partial_x \Phi_{\mu,j})^2 
+ {2\pi \over L} {1\over 4}  \sum_{j} N_{c,j}^2  \;,
\\
H_s & = & 
h  S_0^z + 
\sum_j 
\bigl(h  N_{s,j} + {2\pi \over L}   N_{s,j}^2  
 + J_z {N_{s,j} \over L} S_0^z \bigr)\;.
\label{eq:H_tau}
\end{eqnarray}
From Eqs.~(\ref{eq:shift}), (\ref{2ck:transformationN}), 
(\ref{eq:fermionize}), and (\ref{eq:phase}) immediately follows 
that in the bosonization cut-off scheme simply
\begin{equation}
\delta = {J_z\over 8} \;.
\label{eq:delta(v_z)}
\end{equation}

To prove Eq.~(\ref{eq:ch_imp^glob}) we observe that 
the external field only appears in (\ref{eq:H_tau}). Therefore the 
partition function factorizes as
\begin{eqnarray} 
Z(\beta) & = & Z_0(\beta) \times Z_{s} (\beta,h)\;, 
\nonumber 
\\
Z_s (\beta,h )&  = & \sum_{N_s} \sum_{S_0^z = \pm 1/2} 
e^{-\beta H_s
}
\;.
\nonumber 
\end{eqnarray}
It is easy to evaluate the sum above in the $L\to\infty$ limit 
giving: 
\begin{equation}
Z_s \sim e^{\beta f {L\over 2\pi} {1\over 4} h^2} \times 
{\rm cosh} \Bigl[ {\beta h  \over 2} \bigl(1- f {J_z\over 4\pi} 
\bigr) \Bigr] \;.
\end{equation}
The first term just generates the Pauli susceptibility of a 
free electron gas, while the second corresponds to a free 
spin coupled to a  renormalized magnetic field and 
gives a Curie susceptibility: 
\begin{equation}
\chi_{\rm glob}^{\rm TLS} = {(1- f J_z/4\pi)^2\over 4T}\;.
\end{equation}
Together with Eq.~(\ref{eq:delta(v_z)}), this  yields 
Eq.~(\ref{eq:ch_imp^glob}).

As a further test, one can compare this result with the 
exact relation, Eq.~(\ref{eq:chihighT}) in the small 
coupling limit. Within the Bethe Ansatz cutoff scheme\cite{WiegmanTsvelik}
 $\cos(\mu) = {\cos(J_z/2)/ \cos(J_\perp/2) }$, which in the 
appropriate small coupling limit gives $\mu = J_z/2 + {\cal O} (J_z^2) = 
4 \delta + {\cal O} (\delta^2)$. Substituting this expression into 
Eq.~(\ref{eq:chihighT}) we indeed recover the exact relation 
Eq.~(\ref{eq:ch_imp^glob}) in linear order in $\delta$.


\begin{thebibliography}{99}
\bibitem{TLSreviews} For a review see D.L. Cox and A. Zawadowski, Adv. 
Phys. {\bf 47}, 599 (1998).
\bibitem{VladZaw} K.\ Vlad\'ar and A.\ Zawadowski, 
{ \em Phys.\ Rev. } B {\bf28}, 1564, 1582, 1596 (1983). 
\bibitem{Altshuler_excited}I.L.  Aleiner, B.L. Altshuler, Y.M.
 Galperin, T.A. Shutenko, Phys. Rev. Lett. {\bf 86}, 2629 (2001). 
\bibitem{Smolyarenko} I. Smolyarenko and N.S. Wingreen, 
Phys. Rev. B {\bf 60}, 9675 (1999).
\bibitem{buhrman} D.C.\ Ralph, 
A.W.W.\ Ludwig,  Jan von Delft, and R.A.\ Buhrman, 
Phys.\ Rev.\ Lett.\ {\bf72}, 1064 (1994);
\bibitem{Titanium} 
S.K. Upadhyay, R.N.  Louie, R.A. 
Buhrman, Phys. Rev. B {\bf 56}, 12033 (1997).
\bibitem{ZarJanZaw} 
G. Zar\'and,
Jan von Delft, and A. Zawadowski, Phys. Rev. Lett. {\bf 80}, 1353 (1998).
\bibitem{glazman.90} L. I. Glazman and K. A. Matveev, Sov. Phys. JETP 
{\bf 71}, 1031 (1990).
\bibitem{zarzim}
G. Zar\'and, G.T.  Zim\'anyi, and  F. Wilhelm, 
Phys. Rev. B {\bf 62} 8137 (2000).
\bibitem{dots}
D. Berman, N.B. Zhetinev, and R.C. Ashoori, 
Phys. Rev. Lett. 82, 161 (1999). 
\bibitem{TheoZar} T. A. Costi and G. Zar\'and,  Phys. 
Rev. B {\bf 59}, 12398 (1999).
\bibitem{WiegmanTsvelik} 
A. M. Tsvelik and P. B. Wiegmann, 
{\em Adv. Phys.} {\bf 32}, 453 (1983).
\bibitem{CFT} 
I. Affleck and A. W. W. Ludwig, Phys. Rev. B {\bf 48}, 7297 (1993).
\bibitem{Lowenstein}
J.H. Lowenstein, Phys. Rev. {\bf 29}, 4120 (1984). 
\bibitem{Finkelst} P. B. Vigmann and A. M. Finkel'ste\u{i}n,
Sov. Phys. JETP {\bf 48},102 (1978). 
\bibitem{AndresAndrei} 
N. Andrei and A. Jerez, Phys. Rev. 
Lett. {\bf 74}, 4507 (1995).
\bibitem{andrei.84} N. Andrei and C. Destri, Phys. Rev. Lett. {\bf 52},
364 (1984).
\bibitem{Takahashi}M. Takahashi and M. Suzuki, Prog. Theor. Phys. 
{\bf 48}, 2187 (1972).
\bibitem{TsvelikToulouse}  A. M. Tsvelik, Phys. Rev. B 52, 4366-4370
(1995).
\bibitem{Emeryreview} V. J. Emery and S. A. Kivelson,
in {\em Fundamental Problems in Statistical Mechanics VIII},\/ 
edited by H. van Beijeren
and M. H. Ernst, Elsevier, Amsterdam, (1994).
\bibitem{Andreirev}  N. Andrei, in 
{\it Series on Modern Condensed Matter Physics - Vol. 6}, 
458-551, World Scientific, Lecture Notes of ICTP Summer Course, September
1992. Editors: S. Lundquist, G. Morandi and Yu Lu.
\bibitem{remark} To obtain the prefactor in Eq.~(5) one has to 
study the Toulouse limit of Ref.~\protect{[\onlinecite{TsvelikToulouse}]}.
\bibitem{cbulk}  
$c_{\rm bulk}$ in Eq.~(\protect{\ref{eq:WRglobal}})  denotes 
the {\em total}  specific heat  of the
electron gas, which cannot be obtained from the fused BA equations, 
that capture only contributions from
 the $SU(2)_f$ spin sector (see e.g. Ref.~\protect{\cite{Andrei?}}). 
\bibitem{PangCox} H.-B. Pang and D. L. Cox, Phys. Rev. 
B {\bf 44}, 9454 (1991).
\bibitem{ZarJan}  G. Zar\'and, and J. von Delft,
Phys. Rev. B {\bf 61}, 6918 (2000).
\bibitem{Ye}J. W. Ye, Phys. Rev. Lett. {\bf 77}, 3224 (1996).
\bibitem{VladZimZaw} K. Vlad\'ar, A. Zawadowski 
and G. T. Zim\`anyi, Phys. Rev. B {\bf 37},
2001, 2015 (1988). 
\bibitem{EmeryKivelson} V. J. Emery and S. Kivelson, Phys. Rev. B {\bf 46}, 10\,812 (1992); 
A. M. Sengupta and A. Georges, Phys. Rev. B {\bf 49}, 10020 (1994). 
\bibitem{miracle}
By Eq.~(\ref{eq:connect})
the rescaling of $g$ exactly compensates the rescaling of $h$, and 
therefore the BA equations with   $g = 1$ and $g'=1$ determine 
directly the  impurity free energy.
\bibitem{Jan:boso}  Jan bosonization. 
\bibitem{note-global} Note, however, that the 
corresponding global impurity Wilson ratio for the $f=1$ case 
is independent of both the magnetic field
and the anisotropy, and takes the usual value
$R_{\rm glob}^{\rm imp, f=1}=2$ \cite{WiegmanTsvelik,TheoZar}.
\bibitem{SacramentoSchlottmann} P. D. Sacramento and P. Schlottman, 
Phys. Rev. B {\bf 43}, 13294 (1991).
\bibitem{sassetti.90} M. Sassetti and U. Weiss,
Phys. Rev. Lett. {\bf 65}, 2262 (1990) 
\bibitem{Andrei?} N. Andrei, M. Douglas, 
and A. Jerez, Phys. Rev. B {\bf 58}, 7619 (1998).
\end{thebibliography}
\end{document}